# Field calibration and analysis of a low-cost sensor network based on the existing air quality infrastructure of a city

R. Blaga

Faculty of Physics, West University of Timisoara, V. Parvan 4, 300223 Timisoara, Romania

CONTACT R. Blaga robert.blaga@e-uvt.ro

**ABSTRACT**

Low-cost particulate matter sensors (LCS) are an important source of air quality data, improving the spatial and temporal resolution of data gathered by sparsely placed official monitoring stations. Their readings, however, are subject to bias due to unaccounted for effects pertaining to both the physical properties of the aerosol particles and design limitations of the devices. A calibration model is paramount in order to valorize the output of LCS devices.

In this paper, a calibration model is developed for the LCS network of the municipality of Timişoara, Romania. A regression approach is used for calibrating $PM_{10}$. Several models are tested, considering as independent variables the LCS $PM_{10}$, relative humidity, and temperature. Models with physics-based corrections for relative humidity are found to work best. The calibrated data is tested against data from collocated stations from the National Air Quality Monitoring Network (NAQMN), showing an average performance of nRMSE=7.5µg/m³, R²=0.31.

The calibration model is applied to the city-wide LCS network of the municipality. The yearly average $PM_{10}$ of the network is shown to be similar to that of NAQMN, both being well within the EU yearly $PM_{10}$ standard. For daily $PM_{10}$ values, however, several stations are found that regularly exceed the daily threshold set by the European Union. These violations are not witnessed at the NAQMN stations. This is interpreted as due to the small number of reference stations located in the administrative boundaries of the city, which miss particular localized emission sources that are not collocated with these stations.

## 1. Introduction

PM, short for particulate matter, is a quantity characterizing the mass of suspended particles of various sizes per cubic meter of air. In the context of air pollution studies, PM usually refers to aerosol particles suspended in the atmosphere at ground level. *Fine particles* with dynamic diameters smaller than 2.5µm, denoted $PM_{2.5}$, are especially important for public health studies. Such $PM_{2.5}$ particles can cross the blood-gas barrier in our lungs, contributing to cardiovascular impairments, various metabolic diseases, and adverse effects in infancy (Feng et al. 2016). Larger, *coarse particles* with diameters between 2.5µm and 10µm can also cause great distress to the respiratory system, although they enter the bloodstream to a lesser degree (Lu et al. 2015). Collectively, the mass density of all particles with diameter lower than 10µm—both fine and coarse particles—is denoted $PM_{10}$.

Besides correlation with bad health outcomes, PM is also relevant for solar energy estimation studies. Suspended particles, or aerosols, scatter and absorb part of the incoming solar radiation. In doing so, they influence how much energy there is available to convert into electricity through photovoltaic panels. The magnitude and spatio-temporal variation of this impact is essential for PV operators when performing feasibility studies and operational forecasts for their energy projects (Gutiérrez et al. 2018). For solar energy estimation, the aerosol optical depth (AOD) of the whole atmospheric column is the relevant quantity for estimating this impact (Ruiz-Arias et al. 2016). PM can be linked to AOD if the aerosol composition, hygroscopicity and vertical distribution are known (Tsai et al. 2011). The accurate knowledge of PM and its spatio-temporal distribution at ground level is essential for such studies. The measurement and estimation of PM levels is, thus, of high importance to researchers, public authorities, energy producers, and the general public, alike.

The golden standard of PM measurement is represented by reference devices that measure particle mass through gravimetric methods. Filter-based gravimetric devices capture aerosol particles on a filter and directly measure their mass increment. The prohibitive costs of such reference devices make them accessible only for public authorities or research laboratories. Access of the general public to robust air quality data is usually only possible through the services and reports of the National Environmental Protection Agency (NEPA) in each country.

In this context, ultra-low-cost air quality sensors (LCS) have been becoming ever more popular, with hundreds of such devices now available on the market. Virtually all such devices are based on light-scattering methods. The sensors function by passing a laser beam through a volume of air containing the aerosol particles, and the scattered light is recorded. Through Mie theory, the recorded signal waveform is correlated with the particle scattering cross section. Assuming perfectly spherical particles, the particle volume, and hence the particle diameter, is determined. A further assumption is then usually made about particle density in order to convert particle size into mass density, which is the usual output of the LCS devices. Thus, there are a number of elements where uncertainty is introduced.

The financial accessibility of such sensors has made them a basic tool for citizen scientists who are often frustrated by the opaqueness of their national regulatory agencies. In Romania, where this study is situated, while the agency's datasets are readily available in the public domain, the data has frequent gaps in the hourly PM values. The agency does not make public its exact set of data quality controls, contributing to the suspicion around the reported PM data. "The stations are always turned off exactly when the pollution is highest," is a talking point commonly circulated on social media by active members of civil society, although this is very likely not the case. This lack of confidence of the public in the national authorities comes in the context where Romania has had several infringement procedures initiated against it by the European Commission (EC) for failing to uphold its environmental protection responsibilities (EC 2023), including failing to implement proper industrial air pollution controls (EC 2021) and waste management measures (EC 2017).

While the LCS technology brings crucial accessibility of air quality monitoring to the general public, the interpretation of the PM output from these devices can be misleading. Indeed, the conclusion that emerges from the scientific literature is that these devices can often be unreliable under various scenarios (Karagulian et al. 2019). Several factors which can distort the results of the light-scattering are not accounted for in any way by these devices. As noted, several assumptions and approximations are implicitly made by the manufacturers in order to report PM mass density as output. Uncertainty values or interpretation guidelines are often not provided. Additional confounding factors can also influence the quality of the results. Such factors can pertain both to the physics of the aerosol particles, like the misreading of water droplets for particulate matter, or to the design of the devices, like the inability of the sensors to detect certain categories of particles. A correction of the LCS data is always necessary to obtain accurate PM values. Thus, LCS devices can be an important tool for democracy, but only if they represent reality faithfully. Otherwise, they can just as easily fuel a conspiracy mindset.

Because the LCS errors are based on physical phenomenon, not random error, they can be corrected through parametric calibration equations. Manufacturers, in fact, always calibrate their devices for a range of conditions. However, these calibrations are often performed for a preset type of aerosol in the laboratory. During field deployment, as the aerosol composition differs from the laboratory conditions, the device calibration can lose accuracy (Crilley et al. 2020; Zikova, Hopke, and Ferro 2017; Wang et al. 2021). Furthermore, these calibrations are fixed, not parametric. There is no way to account for the impact of water vapor, for example, in this way. Thus, an in-situ calibration by the user is always necessary in order to obtain accurate PM values.

Calibration studies are performed by collocating the LCS sensor with a reference device. This can happen in the laboratory in the presence of local aerosols (Cavaliere et al. 2018; Liu et al. 2017), or in situ where the sensors are placed (Hong et al. 2021). In situ sometimes implies the deployment of a reference device together with multiple LCS devices in different locations (Hua et al. 2021; Kosmopoulos et al. 2020; Morawska 2018). Such a study is possible for researchers if a fleet of LCS and reference devices is available. However, it is not possible, in general, for the regular citizen to perform such a calibration. Alternatively, lacking a mobile reference device, one could use the existing stations of the national monitoring agency other existing standard devices as reference (Lee et al. 2020; Magi et al. 2020). In this paper we present an example of such a calibration study, adapted for the city of Timișoara, Romania, using the existing sensors placed throughout the city. A cluster of LCS devices approximately collocated with a NAQMN station are used for calibration, which is then applied for the LCS network of the municipality.

This calibration study follows closely the best practices described in Giordano et al. (2021) and Liang (2021), as applied for field deployment (as opposed to laboratory calibration). Of the NAQMN stations from the study location, only one measures $PM_{2.5}$ consistently. The calibration model is thus developed only on $PM_{10}$ data. Furthermore, it is shown in the literature that the calibration of $PM_{2.5}$ can have different subtleties compared to $PM_{10}$ (Jayaratne te al. 2018; Tryner et al. 2020). This makes the direct application of the $PM_{10}$ calibration equations to $PM_{2.5}$ questionable. Therefore, this study focuses exclusively on $PM_{10}$, on daily and yearly timescales.

Several calibration models are tested. The performance of these models is found to be quite low as compared to other calibration studies from the literature, which is interpreted as a limitation of the calibration framework. Despite the relatively low model performance, the study gives an example of what can be achieved using the existing air quality monitoring infrastructure of a city, instead of laboratory equipment. This framework makes the calibration study accessible also for citizen scientists. The best performing model is applied to analyze the daily and yearly $PM_{10}$ levels at the LCS network nodes.

The paper is structured as follows. Section 2 presents the PM sensor networks employed in this study. The working datasets are described in summary, together with the applied quality controls. In Sec. 3 the calibration framework is detailed, and various regression models are tested and validated. The best performing models are applied to the data from the municipal LCS network in order to analyze the city-wide PM levels. Results are summarized and the study is concluded in Sec. 4.

## 2. Particulate matter databases

### 2.1. Second level heading

The measurement of particulate matter mass concentration is paramount for accurate assessment of both health impacts and energetic impacts of ground-level air pollution. In urban environments there are localized sources of particulate matter, like industrial areas and high-traffic roadways, and the aerosol composition can be quite complex. Thus, a high temporal resolution and spatial coverage is necessary to accurately describe the dynamics of PM within a city.

Beginning with the year 2021, the municipality of Timișoara, Romania has developed an extended network of sensors for monitoring the local air quality. The network consists of two sets of low-cost sensors, donated to the municipality by private entities. The devices are placed in a complementary distribution around the city. Additionally, the National Air Quality Monitoring Network (NAQMN) of the Romanian National Environmental Protection Agency (NEPA) has four stations within the administrative boundaries of the city.

Data recorded by the devices from the three networks between 01/10/2021 and 31/09/2023 is considered in this study. The three sensor networks are described first. Then the raw data is described in summary. Finally, a set of pre-processing algorithms is applied to ensure the quality of the working data.

### 2.1.1. ETA2U sensors

The first network consists of sensors provided by the ETA2U foundation (ETA2U 2024) through a partnership with the municipality (ETA2U 2021). The project consists of 27 devices mounted at different locations around the city and the peri-urban region. Both the real-time and historical data can be visualized at www.airdata.ro.

The ETA2U network uses Sensirion SPS3 and Honeywell HPMA115 sensors for measuring PM, and AOSONG AM2320 / Sensirion SHTC1 for temperature / humidity. The sensors were installed in Spring 2021. The producers state a $1000 \mu g/m^3$ maximum range for both devices, with a 10% and 15% uncertainty for the Sensirion and Honeywell sensors, respectively. The lower limit of reliable particle detection is listed as $0.3 \mu m$ by the producers of the Sensirion sensors, while the lower limit of the Honeywell sensors is not provided (Sensirion 2024; Honeywell 2024). The manufacturer gives a 2-year warranty for the devices. The information about which sensors are SPS and which HPMA is not listed on the project website.

### 2.1.2. uRADMonitor sensors

The second municipal sensor fleet consists of devices of a local startup called uRADMonitor (uRADMonitor 2024a), which develops and commercializes the product. uRADMonitor offers the data from all their installed devices in the public domain. The data can be accessed at www.uradmonitor.com. 16 A3 Model devices have been donated by the company to the municipality, and there are also a few private units that are accessible through the company's platform. The donated sensors were mounted in June 2021. As the private devices are not mounted at the same time, their effective age is uncertain, with several being inoperational at the time of this study. Additionally, some of the sensors are mounted on public transport units. In this study only the fixed-point devices are considered, and only the data which spans the study period is retained.

The uRADMonitor devices use Plantower PMS5003 sensors for measuring $PM_1$, $PM_{2.5}$, and $PM_{10}$ (uRADMonitor 2024b). The devices come with a 1-year warranty. The sensors have $1000\mu g/m^3$ maximum and $500\mu g/m^3$ effective range for measuring $PM_{2.5}$. For particles of diameter 0.3 and 0.5$\mu$m the counting efficiency is 50% and 98%. Measurement uncertainty for $PM_{2.5}$ is $\pm10\mu g/m^3$ for mass concentrations up to $100\mu g/m^3$, and $\pm10\%$ above this threshold (Plantower 2024).

The Plantower sensors come with two built-in correction factors (CF=1 and CF=atm), appropriate for different aerosol regimes. The uRADMonitor devices are pre-set to CF=atm. We note, however, that for the low-to-moderate aerosol loading and aerosol composition present at the study location, the difference between the two CFs is, generally, found to be small (Kosmopoulos et al. 2020). Aerosol size bins, also output by the Plantower sensors, are not given by the uRADMonitor devices.

### 2.1.3. Open data platform

In alignment with the EU Directive 2019/1024 regarding open data and the re-use of public sector information, the municipality of Timișoara has developed an open data platform (ODP-TM 2024). The platform gathers environmental, demographic, economic and other types of data from various public sources. The environmental data includes the measurements from the two LCS networks. The data can be accessed directly through an API. The R language code for downloading PM datasets from the two LCS networks is given below. The working dataset used in this study is accessed on January 29, 2023. The uRADMonitor data can be accessed using the following R code (*httr* and *jsonlite* packages are required):

```
#Download uRADM. device IDs:
path <- "https://data.primariatm.ro/api/3/action/datastore_search?
resource_id=735beb67-578d-4486-b472-72bbba7f9981"

res = GET(path)
aux <- fromJSON(rawToChar(res$content))
uRADid <- aux$result$records$id

#Download dataset for each ID:
data <- data.frame()
for (i in 1:length(uRADid)){
  print(uRADid[i])
  path <- paste0("https://data.primariatm.ro/api/3/action/
                 datastore_search_sql?sql=SELECT%20*%20from%20%22d680
                 ddb5-45be-4842-95b6-afdef322991a%22%20WHERE%20%22id%
                 20device%22=%20%27",uRADid[i],"%27")
  res = GET(path)

  data_raw <- fromJSON(rawToChar(res$content))
  data <- rbind(data, data_raw$result$records)
}
```

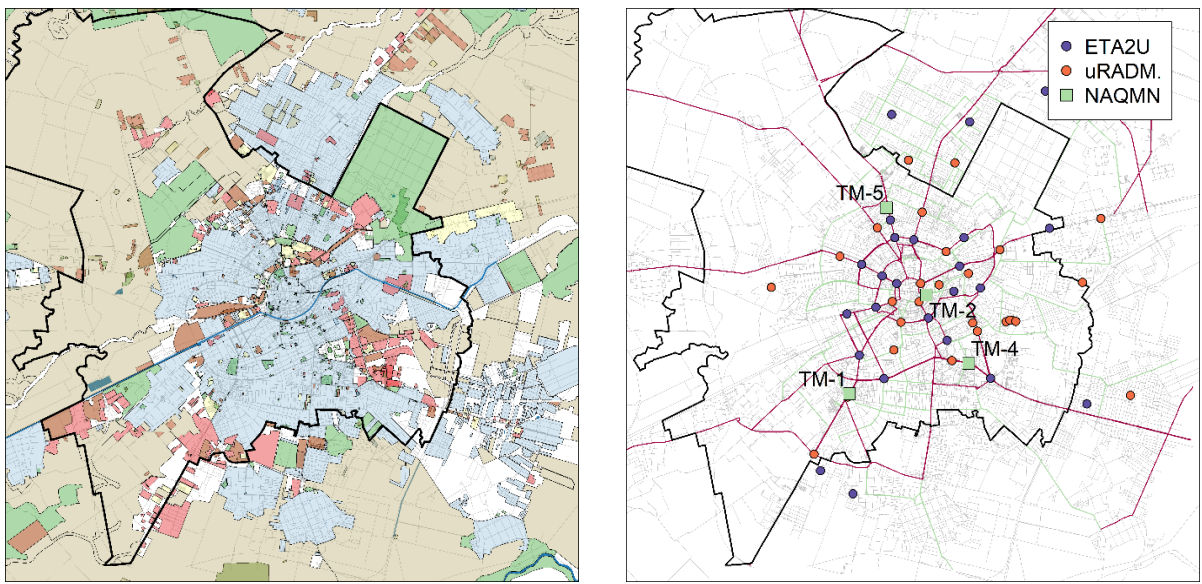

**Fig. 1** Administrative boundary of the municipality of Timișoara, Romania. (a) The major land use types: (blue) residential, (red) industrial; (green) urban green spaces, (dark brown) brownfields, and (light brown) agricultural and other productive land uses. (b) The ETA2U, uRADMonitor and NAQMN sensor networks. The NAQMN station IDs are listed. Major roadways are also shown. The maps are generated using the *osmdata* R package.

The data for the ETA2U network devices can be accessed similarly, by replacing the device IDs, which are "ETA2U-1" through "ETA2U-30".

### 2.1.4. Regulatory agency sensors

The National Air Quality Monitoring Network (NAQMN) was developed in stages starting from 2008. The stations included in this study (see Fig. 1b), have the first recorded gravimetric PM values on January 1, 2009. The reference methods respect the European Standards for the gravimetric measurement of $PM_{10}$ and $PM_{2.5}$ (CS EN 12341). The calibration of the instruments is done separately for the intake system (monthly) and weighing mechanism (yearly). Information regarding the operation of the NAQMN stations was obtained from the National Environmental Protection Agency (NEPA), through Law 544/2001 regarding access to information of public interest. The NAQMN data can be visualized and downloaded from www.calitateaer.ro.

There are four stations of the NAQMN within the administrative boundaries of the city of Timișoara. These have the following device IDs within the NAQMN: TM-1, TM-2, TM-4, and TM-5. The PM measurements of these stations are used as reference in this study.

### 2.2. Data summary

Timișoara is a city located in the western tip of Romania, at 45.749 latitude and 21.227 longitude. With a population above 450.000 in its metropolitan region and a surface area of 130km², it is one of the largest second tier cities after the capital, Bucharest. It is located at an altitude of 90m and has a moderate climate, classified as Cfb in the Geiger-Köppen climate classification system. Summers are hot and dry, with peak temperatures often close to 40ºC and monthly average relative humidity (RH) dropping below 60%, while winters are cold and damp, with monthly temperatures (T) sometimes dipping below freezing point and mean daily relative humidity maintaining close to 100%. The seasonality of temperature and relative humidity is especially important since they strongly impact the readings of low-cost sensors. Fig. 2a shows the RH and T readings from collocated reference and LCS devices, showing a good agreement between the two. Thus, the RH and T measurements from the LCS devices are deemed appropriate to be used as input for calibration in this study.

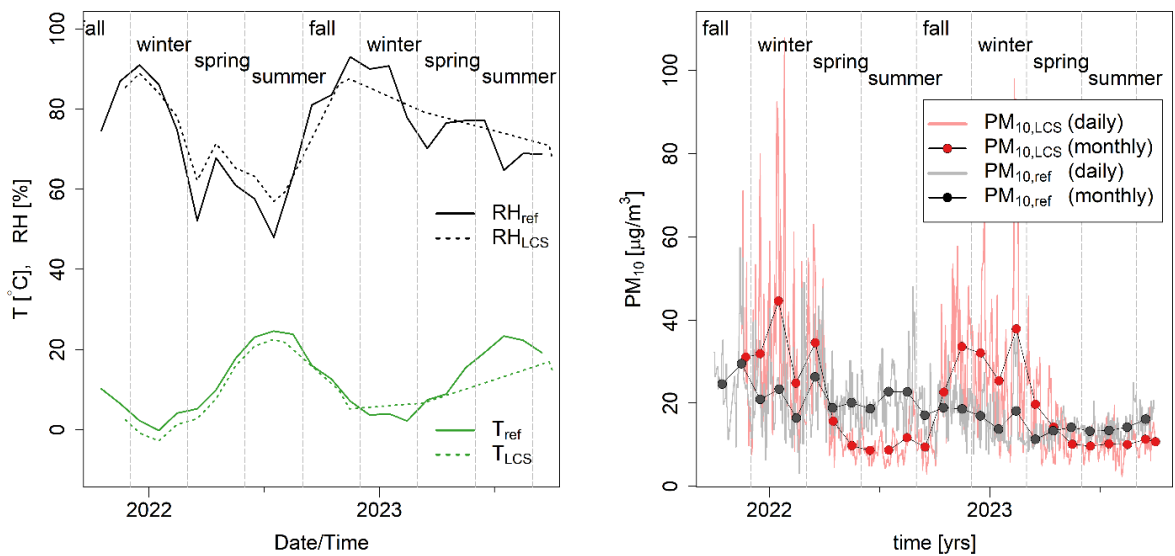

**Fig. 2** (a) Monthly temperature and relative humidity recorded during the study period at a collocated LCS device (82000300) and reference station (TM-4). (b) Daily and monthly $PM_{10}$ recorded with a collocated LCS device (ETA2U-1) and reference station (TM-1).

The National Air Quality Monitoring Network has four stations in the study area. Of the four stations all record $PM_{10}$, but only one (TM-1) records $PM_{2.5}$ regularly. Thus, the focus in this study is exclusively on $PM_{10}$. At the NAQMN stations PM is measured independently both through a gravimetric method and a laser scattering method. Across the stations where both measurement types are available, the difference in mean daily $PM_{10}$ between the two methods is $3.33\mu g/m^3$ and the correlation is 48.3%. Gravimetric methods are considered the golden standard in PM measurements. The relatively low correlation between the two is surprising and casts some doubt on the reliability of the light-scattering data. Hence, the gravimetric data is used here as reference, where possible. At one of the four stations (TM-4) only measurements from the light-scattering method are available. Thus, this station is excluded in the development and validation phase of the calibration model. During the whole study period, the mean daily $PM_{10}$ averaged across the three stations is $21.2\mu g/m^3$. During this time, there is only one day when all three stations record a higher than $50\mu g/m^3$ daily $PM_{10}$. Thus, the city of Timișoara is a place that experiences on average moderate air pollution, with only very rare instances of extremely high $PM_{10}$.

The distribution of the LCS sensors around the city is complementary between the two networks, and they are placed along the most circulated roadways. They cover evenly the central and semi-central areas, but to a lesser degree the more outlying neighborhoods. In terms of land use, their location is mostly in residential areas, with only a sparse placement around industrial zones (see Fig. 1a). In one single instance there are two LCS devices—one from each network—placed very close to each other. In this case a direct comparison of the PM data is possible. The data from the two devices has a correlation of 97.1%. Thus, despite a significant difference in the mean daily $PM_{10}$ measured with these two sensors ($5.8\mu g/m^3$), the data are highly correlated and are, thus, considered inter-comparable—a conclusion which is extrapolated to the whole network henceforth. Fig. 2b shows $PM_{10}$ measurements across the study period from collocated LCS and reference stations. The LCS data shows a very strong seasonality, being sometimes much higher and other times lower than the reference data. Thus, calibration of the LCS data becomes mandatory.

### *2.3. Quality controls*

The raw PM data from the three databases contains many errors and missing values, and hence considerable preprocessing is necessary. Only data from fixed sensors is considered. The Quality Controls (QC) implemented are consistent with the operating guidelines of the LCS sensors. All data with $PM_{10} < -10\mu g/m^3$, $PM_{2.5} < -10\mu g/m^3$ is eliminated. Remaining negative values are set to 0. Similarly, all data with $PM_{10} > 500$ and $PM_{2.5} > 500$ is removed.

Low-cost detectors, have been shown to be quite unreliable in recording particles with diameters between 2.5μm and 10μm, making their $PM_{10}$ readings questionable (Jaffe 2023; Karu and Kelly 2023; Kuula et al. 2020; Ouimette et al. 2022). In particular, both the Plantower and Sensirion sensors used in this study, have been shown to have poor performance during dust-dominated conditions (Kaur and Kelly 2022). The aerosol loading at the locations of the present study is, generally, dominated by fine aerosols, with occasional episodes of wind-blown dust from the Sahara desert. Following Sugimoto et al. (2016), aerosols with a PM fraction $PM_{2.5}/PM_{10} < 0.3$ can be considered as dominated by coarse dust, while at $PM_{2.5}/PM_{10} > 0.8$ the fine particles are dominant. The average PM ratio over our whole working dataset is found to be $PM_{2.5}/PM_{10} = 0.7$, confirming the overwhelming presence of fine aerosol. In order to eliminate the effect of coarse particles and increase the accuracy of our results, all data with $PM_{2.5}/PM_{10} < 0.5$ is removed from the dataset (Kosmopoulos 2020).

Lines containing a missing value for any of the key parameters were also removed: $PM_{10}$, $PM_{2.5}$, temperature, and relative humidity. All QCs together remove 13.45% and 28.23% of the raw ETA2U and uRADMonitor data, respectively. The bulk of the flagged data that was removed is represented by the coarse particles from both networks and the uRADMonitor data from the mobile devices mounted on public transport units, which have all been eliminated. The final databases contain 298607 and 238879 lines, respectively. The NAQMN data is readily available in quality-controlled form and does not require further pre-processing.

When calculating monthly PM averages, a number of 20 days with valid measurements is considered a minimal threshold, while for yearly PM at least 11 monthly values are deemed necessary.

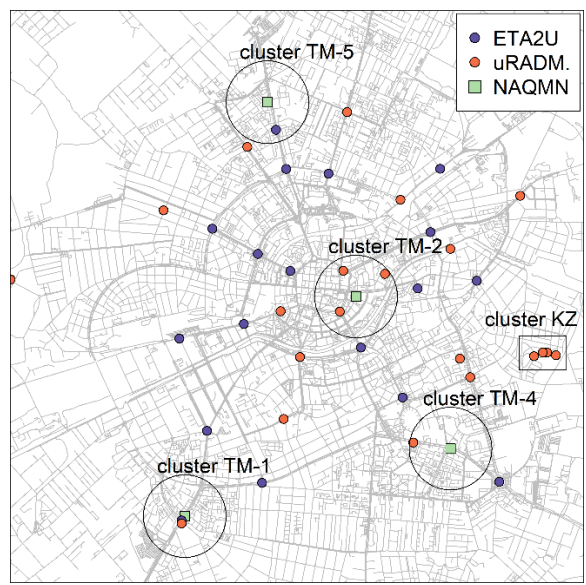

**Fig. 3** The devices used for the development and validation of the calibration model. A 750m radius circle is drawn around each NAQMN station, marking the area where devices are considered collocated for the purpose of field calibration. Numbering of the clusters follows the device ID for each station from the NAQMN. The KZ cluster of outlier LCS devices discussed in Sec. 3.3 is marked with a rectangle.

## 3. Calibration study

In this Section, a model for the calibration of the two LCS networks considered in this study is performed. First, LCS devices placed close to stations of the NAQMN are identified. Then, a set of parametric calibration models are developed and validated. The best performing model combination is then applied for evaluating the city-wide PM$_{10}$ levels, based on the calibrated LCS data.

It must be noted that different sensors generally have different performances, foreclosing a straightforward comparison between their readings. However, the three types of sensors used here—Plantower, Sensirion, and Honeywell—have been found to have comparable accuracy under low-to-moderate aerosol loading, as present at the study site. Nguyen et al. (2021), for example, found that the PMS5003 and SPS30 sensors have 11.3% and 11.4% errors compared to a reference device in the 0-50µg/m³ PM$_{10}$ measurement range. The errors change to 7.6% and 11.4%, respectively, in the 50-100µg/m³ range. The HPM sensors are less studied in the literature and information about their performance is not well established. Bulot et al. (2019) have found that the Honeywell and Plantower sensors show a non-linear relationship, varying with relative humidity. Despite this nonlinearity, the Honeywell readings were found to be closer to the Plantower data than other commercial sensors. Thus, for the purpose of this study, considering the PM levels prevalent at the study locations, the three sensor types are considered commensurate, and their performance is assessed in aggregated form.

### 3.1 *Calibration framework*

#### 3.1.1 *Clustered devices*

To calibrate one instrument in relation to another rigorously, the instruments need to be collocated. Unfortunately, none of the devices from the three networks are strictly collocated. The calibration study is performed here using the LCSs that are closest geographically to each reference system, while being placed in the same environment (e.g., along the same major roadway). Devices located within a 750m radius circular area around each NAQMN station are considered collocated for the purposes of field calibration. Four such clusters are identified, as marked in Fig. 3. The only place where all three types of devices are closely placed is in cluster TM-1. Thus, the LCSs from cluster TM-1 are used for fitting the calibration parameters and the ones from cluster TM-2, TM-4 and TM-5 only for validation. As the analysis focuses on daily (i.e. 24-hour integrated) and yearly PM$_{10}$, the 2-year study interval is deemed appropriate to cover the entire range of meteorological conditions experienced at the study locations (Giordano et al. 2021).

#### 3.1.2 *RH correction and other cofounding factors*

The measurements of LCS generally can be confounded by several factors. Most aerosols are hygroscopic to some degree. Hence, in the presence of moisture they absorb water, growing in volume and mass. This process is called deliquescence. In opposition to most reference systems, the low-cost devices do not have any protocols for removing water from the particles.

All LCS use sensors based on light scattering. The devices pull air into a chamber probed by a laser beam. The angular distribution of the light scattered off the particles that are suspended in the volume of air is used to estimate the scattering cross-section—hence the particle volume—based on Mie theory. Then, the volumetric density is transformed into mass density via built-in transfer functions determined by the developer.

Water can confound the results in several ways. Mie theory is applied with the assumption that the scattering centers are hard spheres, while the aerosol-water solution is non-uniform and not fully opaque to light (Malm et al. 2003). This can lead to an underestimation of the particle volume. Second, the transfer functions are fitted considering particular types of aerosols. This potentially leads to inaccurate estimates for the mass density of the aerosol-water solution, especially considering the complex aerosol mix present in urban environments (Crilley et al. 2020).

For in situ measurements, the hygroscopicity can be determined directly. Such measurements are, however, rare and hard to obtain. In absence of measurements, a simple correction based on relative humidity (RH) is usually performed. The two main approaches in the literature for implementing an RH correction are either a physics-based correction or a linear regression. Currently, there are two main approaches in the literature to the physics-based correction. The first entails a simple RH correction as follows (Giordano et al. 2021):

$$PM_{LCS} = f(RH) \cdot PM_{ref}, \quad f(RH) = r_1 + r_2 \frac{RH}{1-RH^2}, \tag{1}$$

where the parameters $r_1$ and $r_2$ are determined by fitting on data.

The second approach is based on κ-Köhler theory, wherein the hygroscopicity of particles—i.e. their capacity to absorb water from the environment—is quantified as (Di Antonio et al. 2018):

$$d_{dry} = d_{wet}(RH)/g(RH), \quad g(RH) = \left(1 + \kappa \frac{RH}{100-RH}\right)^{1/3}. \tag{2}$$

$d$ is the particle diameter, while the parameter κ—determined by the chemical composition of the particle—characterizes the particle hygroscopicity. Starting from Eq. (2), Crilley et al. (2018) have developed a correction for PM in the form of:

$$PM_{ref} = PM_{LCS}/k(RH), \quad k(RH) = 1 + \frac{\kappa \cdot RH}{1.65(100-RH)}. \tag{3}$$

The second approach to account for the impact of water vapor is based on simple (SLR) or multiple linear regression (MLR). More details about MLR are given in Sec. 3.1.3.

Besides relative humidity, another factor that hinders the direct interpretation of PM data from LCS devices is related to the detection limits of the sensor. Most light-scattering sensors have a lower limit for particle size below which the aerosols are no longer reliably detected. This issue is most commonly called the 'limits of detection' (LOD) in the literature (Liang 2021). For LCS the limit of detection is usually somewhere between 0.1μm and 1μm. The Sensirion and Honeywell sensors from the ETA2U network, as well as the Plantower sensors from the uRADMonitor network, have LOD = 0.3μm, for example. The finite LOD leads to an underestimation of the mass density of the aerosols.

Determination of aerosol size is also a complicated issue in general. Ouimette et al. (2024), for example, have found that the Plantower sensors can undersize the particles in some cases. From the available data it is difficult to determine exactly what effects impact the PM readings in the present study. In any case, it is clear that the LCS underestimate the PM during dry periods. Together all effects that lead to PM underestimation due to aerosol size and/or concentration are called here under the collective term "LOD effect".

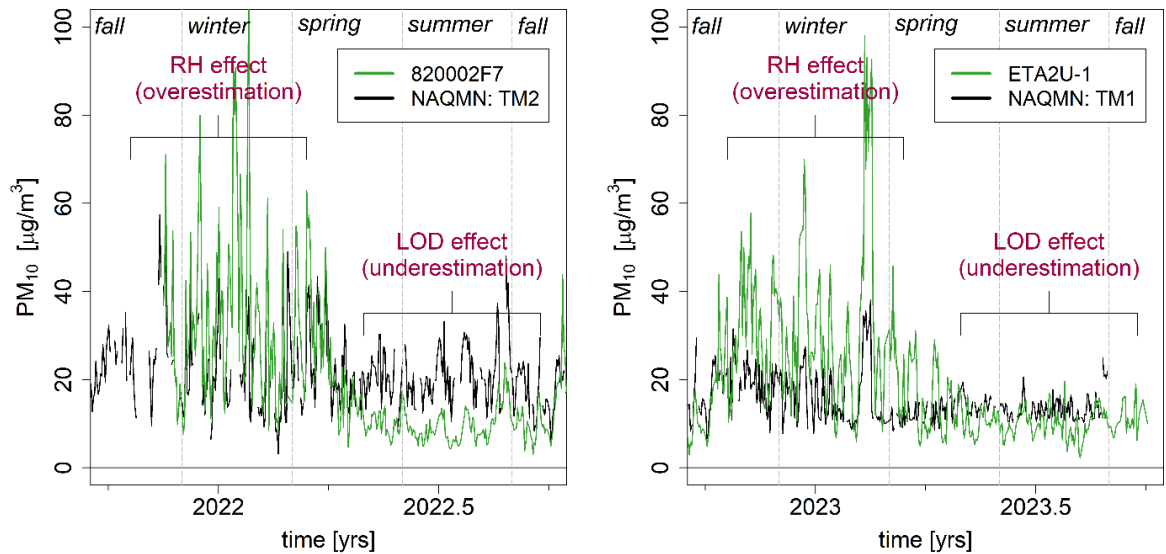

**Fig. 4** $PM_{10}$ measurements from two of the NAQMN stations with their collocated LCS counterparts. The periods when the RH and LOD effects are dominant are highlighted, respectively.

An illustration of both the RH and LOD effects is visible in the comparison between LCS and reference PM measurements, shown in Fig. 4. The plot shows the data from collocated reference and LCS instruments, highlighting the period dominated by each of the two effects. It is clear from Fig. 4 that from mid fall to early spring the RH effect leads to very high PM values recorded with the LCS devices. The LOD effect becomes more prominent from mid spring to early fall, when the aerosols are overall lower in concentration and the fraction of small particles is higher. Note, however, that the LOD effect is present at all times, but is masked by the RH effect when the latter is large.

Other effects also influence the LCS output. For example, because in general the sensors assess only the scattering cross-section in the whole volume, aerosols can mask each other when they are collinear with the scattering beam, leading to a slight underestimation in the aerosol volume (Giordano et al. 2021). The LOD and other smaller effects are more easily captured through the regression approach, because their compound effect is encapsulated in the regression coefficients. Although the physical approach brings the promise of increased precision, all impactful effects would need to be accounted for in order to obtain high accuracy. Thus, in what follows, we focus on regression-based models or combinations of regression and physical corrections.

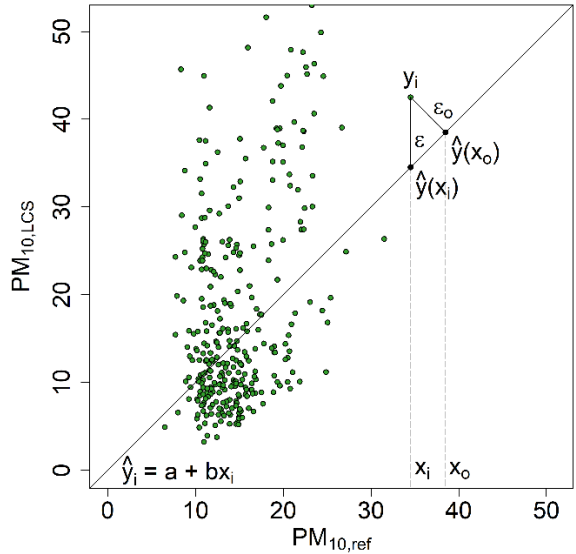

**Fig. 5** Illustration of the differences between LS and LOS routes for optimization of regression coefficients. LS minimizes the average of $\varepsilon^2$, while LOS the average of $\varepsilon_o^2$. Data from devices ETA2U-1 (ETA2U net.) and TM-1 (NAQMN) from the working dataset is shown.

### 3.1.3 RH correction and other cofounding factors

In the MLR approach to PM calibration, the reference $PM_{10}$ is considered as the modelled variable, while the LCS $PM_{10}$, relative humidity (RH), temperature (T), and sometimes wind speed and direction ($\vec{v}$), are considered the explanatory variables. Thus, a general MLR for $PM_{10}$ can be written as:

$$\widehat{PM}_{10,\text{ref}} = a + b \cdot q_1(PM_{10,\text{LCS}}) + c \cdot q_2(RH) + d \cdot q_3(T) + \vec{e} \cdot \vec{q}_4(\vec{v}),$$ (4)

where $q_1(\ )$, $q_2(\ )$, $q_3(\ )$, and $\vec{q}_4(\ )$ are arithmetic functions of the explanatory variables. The regression coefficients $a$, $b$, $c$, $d$, and $\vec{e}$ are then determined by imposing some constraint. The most common is the least squares approach (LS). In LS the average of the square of residuals between measurements and model estimates is minimized with respect to each regression coefficient in part. Other approaches, like Linear Orthogonal Squares (LOS), are sometimes used. In LOS the orthogonal distance from measurement to the modelling line is minimized instead of the residuals (Johnson et al. 2018; Liang 2021). The difference between LS and LOS is illustrated in Fig 5. In this study the LS method is used henceforth.

### 3.2   Calibration of LCS network

In this section, various regression models are tested for the calibration of the LCS network considered in the study. The clustered devices described in Sec. 3.1.1 are used for the development and validation of the calibration model.

#### 3.2.1   Model development

A linear relationship between LCS and reference PM is universally observed in the literature, at least up to moderate-to-high PM values (Han, Symanski, and Stock 2017; Johnson et al. 2018; Liang 2021; Zheng et al. 2018). At very high PM, the LCS measurements lose accuracy, and the relationship is often better modelled with a quadratic or other non-linear correlation (Gao, Cao, and Seto 2015). Considering that the PM values at the measurement locations rarely surpass $100\mu g/m^3$, a linear relationship is assumed henceforth. Furthermore, aerosol composition can affect the readings of low-cost sensors, a problem which potentially requires calibration separately for each aerosol type (He, Kuerbanjiang, and Dhaniyala 2020; Hagan and Kroll 2020; Malings et al. 2020; Sousan et al. 2016; Zou, Clark, and May 2021). In the study area the aerosols are generally urban-industrial, with only a few occasions a year experiencing high desert dust or other inorganic dust dominated conditions. Thus, we assume a relative stability of aerosol composition throughout the study.

In order to find the best calibration model, different approaches for building the model are tested. A general equation of the following form is fitted:

$$\widehat{PM}_{10,\text{ref}} = a + b \cdot q_1(PM_{10,\text{LCS}}) + c \cdot q_2(RH) + d \cdot q_3(T) \tag{5}$$

where $q_1()$, $q_2()$, and $q_3()$ are functions of LCS $PM_{10}[\mu g/m^3]$, relative humidity RH[%] and temperature T[°C], respectively. Wind data is not considered in this study. 3 choices for $q_1$, 5 for $q_2$, and 3 for $q_3$ are tested.

The cluster TM-1 devices (see Sec. 3.1.1) are used for developing the regression model. Of the two LCS devices, the ETA2U sensor is used for calibration (ID: ETA2U-1). In a first test, all models are found to perform quite poorly when applied on the whole dataset. Thus, the dataset is divided into two subsets. The first subset spans the interval from 1st of April to 31st of September, representing hot, relatively dry months (from late Spring to early Autumn in the Northern Hemisphere), while the second subset spans the colder, more humid months from 1st October through 31st March. The two subsets approximately divide the dataset into periods dominated by the RH effect and LOD effect, respectively. All model variants are applied on the two data subsets independently.

The nRMSE of each model version on the testing dataset is shown in Table 1. The last two columns represent model versions in steps, with the physical RH corrections applied first [see Eq. (1) and (3) in Sec. 3.1.2], and a temperature-only regression applied second. The parameters of the f(RH) correction are determined through LS fitting as $r_1 = -0.0476$ and $r_2 = 1.0571$ for the dry months, and $r_1 = 0.2974$ and $r_2 = 1.8032$ for the humid months. The correction k(RH) is determined following the guideline of Svenningsson et al. (2006). The authors recommend a value of $\kappa=0.62$ under polluted environments, containing a mixture of organic and inorganic substances, which is the case here. $RH^0$ and $T^0$ mean that the respective parameter is not included in the regression.

**Table 1** nRMSE of the 21 calibration model variants on the two data subsets.

|         |       | $RH^0$ | $RH^1$ | $RH^2$ | f(RH) | k(RH) | $f^*(RH)$ | $k^*(RH)$ |
|---------|-------|--------|--------|--------|-------|-------|-----------|-----------|
| hot/dry | $T^0$ | 32.56  | 26.71  | 27.08  | 30.05 | 30.12 | 25.44     | 27.82     |
|         | $T^1$ | 30.31  | 26.18  | 26.57  | 29.03 | 29.08 | **24.74** | 26.59     |
| | $T^2$ | 30.32 | 26.30 | 26.66 | 29.04 | 29.09 | 24.88 | 26.73 |
| cold/humid | $T^0$ | 38.68 | 36.21 | 36.36 | 37.41 | 37.43 | 38.78 | 36.04 |
| | $T^1$ | 38.25 | 35.91 | 36.04 | 37.07 | 37.09 | 38.77 | 35.86 |
| | $T^2$ | 38.08 | 35.79 | 35.93 | 36.97 | 36.99 | 38.76 | **35.77** |

Interestingly, the two stepwise applied physical RH correction models work the best. The simple correction [Eq. (1)] works best on the LOD-dominated dataset, while the κ-Köhler theory-based correction [Eq. (3)] works best on the RH effect dominated subset. These results agree with the recommendation from Crilley et al. (2020) to apply κ-Köhler theory for high humidity conditions, and a simple linear correction when RH is below 60%, which roughly conforms with the partition of our dataset. Temperature has an important role during the hot/dry period, while its impact is marginal during the cold/humid months. Thus, the following equation is used as the calibration model:

$$\widehat{PM}_{10,ref} = \begin{cases} 8.2392 + 0.6066\frac{PM_{10,LCS}}{f(RH)} + 0.1662T, & \text{hot/dry months} \\ 9.8838 + 0.6798\frac{PM_{10,LCS}}{k(RH)} + 0.01028T^2, & \text{cold/humid months} \end{cases} \quad (6)$$

Note that the intercept was not set to zero as would be required physically. This is justified by the LOD problem as well as the high sensor relative uncertainty at low PM. Indeed, model versions with vanishing intercept perform significantly worse than the models in Table 1, over the testing dataset.

### 3.2.2  Model validation

The calibration model Eq. (6) is validated against data from clusters TM-1, TM-2, and TM-5, as defined in Sec. 3.1.1. The results are listed in Table 2. The models achieve mean bias within 9-17% and root mean square error between 33-35%. These values are within the error margin of the devices. Indeed the 1.7-4.6μg/m³ value range for MBE and 6.3-8.7μg/m³ for RMSE, is below the 10μg/m³ sensor uncertainty (see Sec. 2.1). Thus, the calibration is deemed reliable to use for the analysis of the municipal PM levels based on data recorded by the two LCS networks.

In interpreting the results, one also has to take account of the issue of consistency between model optimization strategy and intended use. Discussed by various authors in the field of forecast models (Yang et al. 2020), consistency is an important, often neglected, issue also for estimation models. The *lm* function in R uses the QR decomposition method to solve the LS equations, to obtain the best fit. This leads to a low nRMSE, but relatively poor nMBE and $R^2$. nMBE can be improved by a simple linear calibration, but that would lead to increased nRMSE. A particular optimization strategy always represents a trade-off of accuracy in one indicator versus another. Thus, it should be clear which error measure one is interested in optimizing, based on the studied problem.

In the case of PM modelling, different problems are of interest. Air quality guidelines are usually given in terms of yearly and daily PM. The yearly threshold is absolute, i.e. the measured PM values are either below or above it. For the daily threshold the number of days per year when it is surpassed is relevant, meaning the threshold is interpreted in a statistical way. Furthermore, sometimes not a statistical but a time series perspective of PM is relevant, like when analyzing an episode of urban pollution or dust storm.

The mean bias error (MBE) captures the model bias, and, thus, it is best to be optimized when yearly PM is of interest. $R^2$ captures the degree to which variations in the dependent variable ($PM_{ref}$) are explained by the explanatory variable ($PM_{LCS}$). Thus, optimizing for $R^2$ might be useful when daily PM values are of interest. RMSE taxes single large modelling errors, making it appropriate for time series analysis.

**Table 2** Validation of the calibration model Eq. (6) on the data from clusters TM-1, TM-2, TM-5.

| Cluster | Network | Id | MBE [µg/m³] | nMBE [%] | RMSE [µg/m³] | nRMSE [%] | $R^2$ | $<PM_{10,ref}>$ [µg/m³] | $<PM_{10,LCS}>$ [µg/m³] |
|---|---|---|---|---|---|---|---|---|---|
| 1 | uRADM. | 82000302 | 3.46 | 16.73 | 7.24 | 35.05 | 0.17 | 20.66 | 17.20 |
| 2 | uRADM. | 820002F6 | 3.21 | 14.09 | 7.77 | 34.14 | 0.36 | 22.76 | 19.55 |
| 2 | uRADM. | 820002F7 | 1.73 | 9.33 | 6.27 | 33.89 | 0.41 | 18.50 | 16.77 |
| 5 | ETA2U | ETA2U-6 | 4.6 | 17.18 | 8.72 | 32.58 | 0.29 | 26.77 | 22.17 |

Finally, it must be noted that $R^2$ values are very low compared to other calibration studies from the literature. We interpret these values as arising mainly from two factors. First, the study location experiences low-to-moderate pollution levels. It is known that LCS sensors strongly underperform in low pollution environments. Johnson et al. (2018), for example, found that a generally well performing LCS sensor, with $R^2$>0.8, had very poor performance when the $PM_{2.5}$ was restricted to below 40µg/m³ ($R^2$~0) [47]. Liang (2021) notes that this is due to the "LCSs' relatively high limit of detection and noisier response", in term due to "the composition of much smaller particles [being] higher in the low concentration air mix". In our case the low correlation is found to persist despite the calibration.

The second important factor is the placement of the reference and LCS devices within a cluster, which are not strictly co-located. The clusters have a diameter of 1.5km which translates into different $PM_{10}$ readings at different devices. In the case of cluster 2, for example, the reference device is off the main road, shielded by a group of tall buildings, while the one in cluster 5 is placed on urban green space near a primary road. Meanwhile, the LCS devices are all placed strictly along the primary roads.

Even so, the uRADMonitor device from cluster 1 achieves a particularly small $R^2$, while the MBE and RMSE are not outliers. It is possible that this device experiences some drift or other technical problem. In any case, the reported $R^2$ and other values are well outside the performance guidelines of the US EPA [61]. Thus, the LCS readings analyzed in this study should be considered only as a qualitative indication, not a rigorous quantitative description of air quality.

### 3.3 Calibration of LCS network

The calibration model Eq. (6), with the additional bias removal described at the end of Sec. 3.2.2, is applied to all LCS devices with valid data from the two networks. Overall, the calibrated LCS PM data is commensurate with the NAQMN data. After calibration, the mean yearly $PM_{10}$ during the year 2022 across all the LCS devices is 25.8µg/m³, compared to 21.24µg/m³ for the three NAQMN stations. Thus, we conclude that the NAQMN data, while slightly underestimating, does provide a reasonably accurate picture of average yearly $PM_{10}$ exposure across the city.

However, when looking at individual sensor data, it is found that the NAQMN, due to the small number of stations, misses particular localized emission sources. The standard deviation of all the LCS devices with valid data is σ=7.93 µg/m³. The two LCS devices that have values one standard deviation above the city-wide mean are both located in the same neighborhood, part of cluster KZ as marked in Fig. 3. The area in question—the Kuncz neighborhood—is one of the comparatively poor neighborhoods of the city. The street with the households where the devices are mounted is unpaved and partially lacks basic services. Heating is done through wood burning stoves, which are deemed the likely culprit for the high recorded $PM_{10}$ values. The personal exposure of the nearby inhabitants is impossible to estimate through any kind of modelling based purely on the NAQMN data.

The EU air quality standards (EU AQS) specify a 50µg/m³ threshold for daily averaged $PM_{10}$, which can be exceeded at most 35 times a year. During the year 2022 the NAQMN stations exceed the threshold at most 2 times, which is well within the EU standard. In contrast, the LCS devices on average cross the threshold 37 times. The two devices from the Kuncz neighborhood that record the highest annual $PM_{10}$, cross the daily threshold a staggering 64 and 164 times. One of the two devices also exceeds the yearly 40µg/m³ threshold. Even accounting for the uncertainty in the LCS data, these values far exceed the air quality guidelines set by the EU. Indeed all four devices, marked as cluster KZ in Fig. 3, cross the daily threshold more than 60 times, although for the others yearly values cannot be computed due to months with missing values. Thus, the pollution levels do not pertain to one individual household but are representative of the neighborhood.

The overall conclusion from this study is that the stations of the National Air Quality Monitoring Network are adequate for capturing the average yearly $PM_{10}$ values at the city level. On the other hand, these stations fail to capture emissions from particular localized sources, leading to a severe underestimation of individual daily exposures of citizens in some cases.

## 4. Conclusion

In this study, a calibration model was developed for the low-cost air quality sensor network of the municipality of Timișoara, Romania. A field calibration setup was designed by identifying LCS devices closely located to the stations of the national regulatory agency, which were taken as reference. Four such clusters were identified. One cluster was used for model development and the others for validation. Several regression models for calibrating the $PM_{10}$ measurements from the LCS devices were tested. The dataset was divided into two, into a period with hotter, dryer weather and a period with colder, more humid weather, and the models were fitted separately on each.

The best models that were found used a physical relative humidity correction applied to the LCS $PM_{10}$, and then a regression model with the corrected PM and temperature as explanatory variables. A mean bias across all models varying between 1.7-4.6μg/m³ was found, with a RMSE between 6.3-8.7μg/m³. The errors are all within sensor uncertainty. The necessity of a further bias correction was found when evaluating yearly $PM_{10}$.

The best calibration model was applied to the entire fleet of LCS devices located within the administrative boundaries of the city. In the first instance, the LCS measurements confirmed that the stations of the regulatory agency capture the mean yearly $PM_{10}$ experienced throughout the city reasonably well, having only a slight underestimation, despite their sparse placement. The values comply with EU air quality regulation. In a deeper analysis, however, it was found that some localized sources were missed by the reference devices. In particular, two sensors were identified which consistently violated the air quality standards imposed by the EU, surpassing the 50μg/m³ daily $PM_{10}$ threshold 71 and 172 times, respectively.

The neighborhood where these sensors are located is relatively poor, and it is common that citizens heat their homes using wood burning stoves. It is speculated that this factor leads to the consistently high measured $PM_{10}$ values. The very high personal exposure of the citizens of this neighborhood is impossible to extrapolate based purely on the data of the regulatory agency. Thus, a high spatio-temporal resolution for the city-wide PM levels is deemed mandatory for obtaining a full picture of air quality.

## Acknowledgement

We thank the following persons: C. Bleotu, for his invaluable work of maintaining the Open Data Platform of Timișoara and offering assistance for interacting with the data from the portal; R. Motisan from uRADmonitor, and I. Tepeneu from ETA2U foundation, for offering further information on their respective sensor networks; the people at the Timișoara Town Hall who worked so that the data used in this study is available in the public domain, and easily accesible.

## References

Bulot, F.M., S.J. Johnston, P.J. Basford, N.H. Easton, M. Apetroaie-Cristea, G.L. Foster, A.K. Morris, S.J. Cox, and M. Loxham. 2019. Long-term field comparison of multiple low-cost particulate matter sensors in an outdoor urban environment. *Scientific reports* 9: 7497.

Cavaliere, A., F. Carotenuto, F. Di Gennaro, B. Gioli, G. Gualtieri, F. Martelli, A. Matese, P. Toscano, C. Vagnoli, and A. Zaldei. 2018. Development of low-cost air quality stations for next generation monitoring networks: Calibration and validation of PM2.5 and PM10 sensors. *Sensors* 18: 2843. https://doi.org/10.3390/s18092843

Crilley, L.R., M. Shaw, R. Pound, L.J. Kramer, R. Price, S. Young, A.C. Lewis, and F.D. Pope. 2018. Evaluation of a low-cost optical particle counter (Alphasense OPC-N2) for ambient air monitoring. *Atmospheric Measurement Techniques* 11: 709-20. https://doi.org/10.5194/amt-11-709-2018

Crilley, L.R., A. Singh, L.J. Kramer, M.D. Shaw, M.S. Alam, J.S. Apte, W.J. Bloss, L. Hildebrandt Ruiz, P. Fu, W. Fu, and S. Gani. 2020. Effect of aerosol composition on the performance of low-cost optical particle counter correction factors. *Atmospheric Measurement Techniques* 13: 1181-93. https://doi.org/10.5194/amt-13-1181-2020

CSN EN 12341. Ambient air - Standard gravimetric measurement method for the determination of the PM10 or PM2,5 mass concentration of suspended particulate matter. Accessed December 12, 2023. https://www.en-standard.eu/csn-en-12341-ambient-air-standard-gravimetric-measurement-method-for-the-determination-of-the-pm10-or-pm2-5-mass-concentration-of-suspended-particulate-matter/

Di Antonio, A., O.A. Popoola, B. Ouyang, J. Saffell, and R.L. Jones. 2018. Developing a relative humidity correction for low-cost sensors measuring ambient particulate matter. *Sensors* 18: 2790. https://doi.org/10.3390/s18092790

Duvall, R., A. Clements, G. Hagler, A. Kamal, V. Kilaru, L. Goodman, S. Frederick, K. Barkjohn, I. VonWald, D. Greene, and T. Dye. 2021. Performance Testing Protocols, Metrics, and Target Values for Fine Particulate Matter Air Sensors: Use in Ambient, Outdoor, Fixed Sites, Non-Regulatory Supplemental and Informational Monitoring Applications. *US EPA Office of Research and Development*.

ETA2U 2021. Partnership with the municipality of Timișoara, for mounting air quality sensors throughout the city 2021. Accessed February 9, 2024. https://www.airdata.ro/info

ETA2U 2024. ETA2U Foundation website: www.fundatia-eta2u.ro

EU Air Quality Standards. Accessed February 12, 2024. https://environment.ec.europa.eu/topics/air/air-quality/eu-air-quality-standards_en

EU Directive 2019/2014. Accessed November 12, 2023. https://eur-lex.europa.eu/eli/dir/2019/1024/oj.

European Comission 2017. Waste: Commission refers Romania to Court of Justice for failing to adopt national measures on waste management and waste prevention. Accessed January 30, 2024. https://ec.europa.eu/commission/presscorner/detail/ES/IP_17_1047

European Comission 2021. Air quality: Commission decides to refer Romania to the Court of Justice of the European Union for failure to comply with EU clean air and industrial emissions legislation. Accessed January 30, 2024. https://ec.europa.eu/commission/presscorner/detail/en/ip_21_6264

European Comission 2024. Key decisions of November 2023 infringement package. Accessed January 30, 2024. https://ec.europa.eu/commission/presscorner/detail/en/inf_23_5380

Feng, S., D. Gao, F. Liao, F. Zhou, and X. Wang. 2016. The health effects of ambient PM2. 5 and potential mechanisms. *Ecotoxicology and environmental safety* 128: 67-74. https://doi.org/10.1016/j.ecoenv.2016.01.030

Gao, M., J. Cao, and E. Seto. 2015. A distributed network of low-cost continuous reading sensors to measure spatiotemporal variations of PM2.5 in Xi'an, China. *Environmental Pollution* 199: 56-65. https://doi.org/10.1016/j.envpol.2015.01.013

Giordano, M.R., C. Malings, S.N. Pandis, A.A. Presto, V.F. McNeill, D.M. Westervelt, M. Beekmann, and R. Subramanian. 2021. From low-cost sensors to high-quality data: A summary of challenges and best practices for effectively calibrating low-cost particulate matter mass sensors. *Journal of Aerosol Science* 158: 105833. https://doi.org/10.1016/j.jaerosci.2021.105833

Gutiérrez, C., S. Somot, P. Nabat, M. Mallet, M.Á. Gaertner, and O. Perpiñán. 2018. Impact of aerosols on the spatiotemporal variability of photovoltaic energy production in the Euro-Mediterranean area. *Solar Energy* 174: 1142-1152. https://doi.org/10.1016/j.solener.2018.09.085

Hagan, D.H. and J.H. Kroll. 2020. Assessing the accuracy of low-cost optical particle sensors using a physics-based approach. *Atmospheric Measurement Techniques* 13: 6343-55. https://doi.org/10.5194/amt-13-6343-2020

Han, I., E. Symanski, and T.H. Stock. 2017. Feasibility of using low-cost portable particle monitors for measurement of fine and coarse particulate matter in urban ambient air. *Journal of the Air & Waste Management Association* 67: 330-40. https://doi.org/10.1080/10962247.2016.1241195

He, M., N. Kuerbanjiang, and S. Dhaniyala. 2020. Performance characteristics of the low-cost Plantower PMS optical sensor. *Aerosol Science and Technology* 54: 232-41. https://doi.org/10.1080/02786826.2019.1696015

Honeywell 2024. Honeywell Particulate Matter Sensors. Accessed December 12, 2023. https://sps.honeywell.com/us/en/products/advanced-sensing-technologies/healthcare-sensing/particulate-matter-sensors/hpm-series#specifications

Hong, G.H., T.C. Le, J.W. Tu, C. Wang, S.C. Chang, J.Y. Yu, G.Y. Lin, S.G. Aggarwal, and C.J. Tsai. 2021. Long-term evaluation and calibration of three types of low-cost PM2. 5 sensors at different air quality monitoring stations. *Journal of Aerosol Science* 157: 105829. https://doi.org/10.1016/j.jaerosci.2021.105829

Hua, J., Y. Zhang, B. de Foy, X. Mei, J. Shang, Y. Zhang, I.D. Sulaymon, and D. Zhou. 2021. Improved PM2. 5 concentration estimates from low-cost sensors using calibration models categorized by relative humidity. *Aerosol Science and Technology* 55: 600-13. https://doi.org/10.1080/02786826.2021.1873911

Jaffe, D.A., K. Thompson, B. Finley, M. Nelson, J. Ouimette, and E. Andrews. 2023. An evaluation of the US EPA's correction equation for PurpleAir sensor data in smoke, dust, and wintertime urban pollution events. *Atmospheric Measurement Techniques* 16: 1311-22.

Jayaratne, R., X. Liu, P. Thai, M. Dunbabin, and L. Morawska. 2018. The influence of humidity on the performance of a low-cost air particle mass sensor and the effect of atmospheric fog. *Atmospheric Measurement Techniques* 11: 4883-90. https://doi.org/10.5194/amt-11-4883-2018

Johnson, K.K., M.H. Bergin, A.G. Russell, and G.S. Hagler. 2018. Field test of several low-cost particulate matter sensors in high and low concentration urban environments. *Aerosol and Air Quality Research* 18: 565. https://doi.org/10.4209/aaqr.2017.10.0418

Karagulian, F., M. Barbiere, A. Kotsev, L. Spinelle, M. Gerboles, F. Lagler, N. Redon, S. Crunaire, and A. Borowiak. 2019. Review of the performance of low-cost sensors for air quality monitoring. *Atmosphere* 10: 506. https://doi.org/10.3390/atmos10090506

Kaur, K. and K.E. Kelly. 2022. Performance evaluation of the Alphasense OPC-N3 and Plantower PMS5003 sensor in measuring dust events in the Salt Lake Valley, Utah. *Atmospheric Measurement Techniques Discussions* 2022: 1-27.

Kaur, K. and K.E. Kelly. 2023. Laboratory evaluation of the Alphasense OPC-N3, and the Plantower PMS5003 and PMS6003 sensors. *Journal of Aerosol Science* 171: 106181.

Kosmopoulos, G., V. Salamalikis, S.N. Pandis, P. Yannopoulos, A.A. Bloutsos, and A. Kazantzidis. 2020. Low-cost sensors for measuring airborne particulate matter: Field evaluation and calibration at a South-Eastern European site. *Science of The Total Environment* 748: 141396. https://doi.org/10.1016/j.scitotenv.2020.141396

Kuula, J., T. Mäkelä, M. Aurela, K. Teinilä, S. Varjonen, Ó. González, and H. Timonen. 2020. Laboratory evaluation of particle-size selectivity of optical low-cost particulate matter sensors. *Atmospheric Measurement Techniques* 13: 2413-23.

Law 544/2001. Accessed December 22, 2023. https://legislatie.just.ro/Public/DetaliiDocument/31413.

Lee, H., J. Kang, S. Kim, Y. Im, S. Yoo, and D. Lee. 2020. Long-term evaluation and calibration of low-cost particulate matter (PM) sensor. *Sensors* 20: 3617. https://doi.org/10.3390/s20133617

Liang, L. 2021. Calibrating low-cost sensors for ambient air monitoring: Techniques, trends, and challenges. *Environmental Research* 197: 111163. https://doi.org/10.1016/j.envres.2021.111163

Liu, D., Q. Zhang, J. Jiang, and D.R. Chen. 2017. Performance calibration of low-cost and portable particular matter (PM) sensors. *Journal of Aerosol Science* 112: 1-10. https://doi.org/10.1016/j.jaerosci.2017.05.011

Lu, F., D. Xu, Y. Cheng, S. Dong, C. Guo, X. Jiang, and X. Zheng. 2015. Systematic review and meta-analysis of the adverse health effects of ambient PM2.5 and PM10 pollution in the Chinese population. *Environmental research* 136: 196-204. https://doi.org/10.1016/j.envres.2014.06.029

Malings, C., R. Tanzer, A. Hauryliuk, P.K. Saha, A.L. Robinson, A.A. Presto, and R. Subramanian. 2020. Fine particle mass monitoring with low-cost sensors: Corrections and long-term performance evaluation. *Aerosol Science and Technology* 54: 160-74. https://doi.org/10.1080/02786826.2019.1623863

Magi, B.I., C. Cupini, J. Francis, M. Green, and C. Hauser. 2020. Evaluation of PM2. 5 measured in an urban setting using a low-cost optical particle counter and a Federal Equivalent Method Beta Attenuation Monitor. *Aerosol Science and Technology* 54: 147-59. https://doi.org/10.1080/02786826.2019.1619915

Malm, W.C., D.E. Day, S.M. Kreidenweis, J.L. Collett, and T. Lee. 2003. Humidity-dependent optical properties of fine particles during the Big Bend Regional Aerosol and Visibility Observational Study. *Journal of Geophysical Research: Atmospheres* 108. https://doi.org/10.1029/2002JD002998

Morawska, L., P.K. Thai, X. Liu, A. Asumadu-Sakyi, G. Ayoko, A. Bartonova, A. Bedini, F. Chai, B. Christensen, M. Dunbabin, and J. Gao. 2018. Applications of low-cost sensing technologies for air quality monitoring and exposure assessment: How far have they gone? *Environment International* 116: 286-99. https://doi.org/10.1016/j.envint.2018.04.018

National Air Quality Monitoring Network (NAQMN) [Rețeaua Națională de Monitorizare a Calității Aerului (RNMCA)] 2024. Accessed December 22, 2023. https://www.calitateaer.ro/.

National Environmental Protection Agency (NEPA) [Agenția Națională de Protecția Mediului (ANPM)]. Accessed December 22, 2023. http://www.anpm.ro/

Nguyen, N.H., H.X. Nguyen, T.T. Le, and C.D. Vu. 2021. Evaluating low-cost commercially available sensors for air quality monitoring and application of sensor calibration methods for improving accuracy. *Open Journal of Air Pollution* 10: 1.

ODP-TM 2024. Open data platform of the municipality of Timișoara. Accessed January 31, 2023. https://data.primariatm.ro/

Ouimette, J.R., W.C. Malm, B.A. Schichtel, P.J. Sheridan, E. Andrews, J.A. Ogren, and W.P. Arnott. 2022. Evaluating the PurpleAir monitor as an aerosol light scattering instrument. *Atmospheric Measurement Techniques* 15: 655-76.

Ouimette, J., W.P. Arnott, P. Laven, R. Whitwell, N. Radhakrishnan, S. Dhaniyala, M. Sandink, J. Tryner, and J. Volckens. 2024. Fundamentals of low-cost aerosol sensor design and operation. *Aerosol Science and Technology* 58: 1-15.

Plantower 2024. Plantower PMS5003 sensor. Accessed February 12, 2024. https://www.plantower.com/en/products_33/74.html

Ruiz-Arias, J.A., C.A. Gueymard, F.J. Santos-Alamillos, and D. Pozo-Vázquez. 2016. Worldwide impact of aerosol's time scale on the predicted long-term concentrating solar power potential. *Scientific Reports* 6: 30546. https://doi.org/10.1038/srep30546

Sensirion 2024. Sensirion Product Catalog. Accessed December 15, 2023. https://www.sensirion.com/products/catalog/?category=Particulate%20matter

Sousan, S., K. Koehler, G. Thomas, J.H. Park, M. Hillman, A. Halterman, and T.M. Peters. 2016. Inter-comparison of low-cost sensors for measuring the mass concentration of occupational aerosols. *Aerosol Science and Technology* 50: 462-73. http://dx.doi.org/10.1080/02786826.2016.1162901

Sugimoto, N., A. Shimizu, I. Matsui, and M. Nishikawa. 2016. A method for estimating the fraction of mineral dust in particulate matter using PM2. 5-to-PM10 ratios. *Particuology* 28: 114-20.

Svenningsson, B., J. Rissler, E. Swietlicki, M. Mircea, M. Bilde, M.C. Facchini, S. Decesari, S. Fuzzi, J. Zhou, J. Mønster, and T. Rosenørn. 2006. Hygroscopic growth and critical supersaturations for mixed aerosol particles of inorganic and organic compounds of atmospheric relevance. *Atmospheric Chemistry and Physics* 6: 1937-52.

Tryner, J., J. Mehaffy, D. Miller-Lionberg, and J. Volckens. 2020. Effects of aerosol type and simulated aging on performance of low-cost PM sensors. *Journal of Aerosol Science* 150: 105654. https://doi.org/10.1016/j.jaerosci.2020.105654

Tsai, T.C., Y.J. Jeng, D.A. Chu, J.P. Chen, and S.C. Chang. 2011. Analysis of the relationship between MODIS aerosol optical depth and particulate matter from 2006 to 2008. *Atmospheric Environment* 45: 4777-88. https://doi.org/10.1016/j.atmosenv.2009.10.006

uRADMonitor 2024a. uRADMonitor website: https://www.uradmonitor.com/

uRADMonitor 2024b. Model A3 product. Accessed December 15, 2023. https://www.uradmonitor.com/products/#4

Wang, P., F. Xu, H. Gui, H. Wang, and D.R. Chen. 2021. Effect of relative humidity on the performance of five cost-effective PM sensors. *Aerosol Science and Technology* 55: 957-74. https://doi.org/10.1080/02786826.2021.1910136

Yang, D., S. Alessandrini, J. Antonanzas, F. Antonanzas-Torres, V. Badescu, H.G. Beyer, R. Blaga, J. Boland, J.M. Bright, C.F. Coimbra, and M. David. 2020. Verification of deterministic solar forecasts. *Solar Energy* 210: 20-37. https://doi.org/10.1016/j.solener.2020.04.019

Zheng, T., M.H. Bergin, K.K. Johnson, S.N. Tripathi, S. Shirodkar, M.S. Landis, R. Sutaria, and D.E. Carlson. 2018. Field evaluation of low-cost particulate matter sensors in high-and low-concentration environments. *Atmospheric Measurement Techniques* 11: 4823-46. https://doi.org/10.5194/amt-11-4823-2018

Zikova, N., P.K. Hopke, and A.R. Ferro. 2017. Evaluation of new low-cost particle monitors for PM2.5 concentrations measurements. *Journal of Aerosol Science* 105: 24-34. https://doi.org/10.1016/j.jaerosci.2016.11.010

Zou, Y., J.D. Clark, and A.A. May. 2021. Laboratory evaluation of the effects of particle size and composition on the performance of integrated devices containing Plantower particle sensors. *Aerosol Science and Technology* 55: 848-58. https://doi.org/10.1080/02786826.2021.1905148